%
%


\magnification=\magstep0
\nopagenumbers
\parindent 25 pt
\parskip = 1 mm plus 0.3mm

\vsize = 23.80 cm
\hsize = 16.54 cm
\voffset = 0.0 mm
\hoffset = 0.0 mm
\tolerance=1200
\hyphenpenalty=1000
\newtoks\footline \footline={\hfil}
\vglue 0.05in

\def\sectionhead#1{\sectionskip\centerline{\twelvebf #1}\vskip6pt}
\def\sectionskip{\penalty-200\vskip24pt plus12pt minus6pt}
\def\delt{\scriptstyle \Delta }

\def\eol     {\hfil\break}
\def\endpage {\vfill\supereject}


\medmuskip=4mu minus1mu
\thickmuskip=5mu minus1mu
\newcount\liczbamarg
\newcount\nrmarg
\newdimen\lmd
\newdimen\lml
\newdimen\ldd
\newdimen\ldl
\lmd=0.05ex
\lml=0.31em
\ldd=0.12ex
\ldl=0.68em


\font\sixrm=cmr6
\font\sixi=cmmi6
\font\sixsy=cmsy6

\font\sevenrm=cmr7
\font\seveni=cmmi7
\font\sevensy=cmsy7

\font\tenrm=cmr10
\font\teni=cmmi10
\font\tensy=cmsy10
\font\tenit=cmti10
\font\tensl=cmsl10
\font\tenbf=cmbx10
\font\tentt=cmtt10

\font\twelverm=cmr10  scaled \magstep1
\font\twelvei=cmmi10  scaled \magstep1
\font\twelvesy=cmsy10 scaled \magstep1
\font\twelveit=cmti10 scaled \magstep1
\font\twelvesl=cmsl10 scaled \magstep1
\font\twelvebf=cmbx10 scaled \magstep1
\font\twelvett=cmtt10 scaled \magstep1

\font\fourteenbf=cmbx10 scaled \magstep2


\def\tenpoint{%
\def\rm{\fam0\tenrm}%
\def\it{\fam\itfam\tenit}%
\def\sl{\fam\slfam\tensl}%
\def\bf{\fam\bffam\tenbf}%
\def\tt{\fam\ttfam\tentt}%
\textfont0=\tenrm \scriptfont0=\sevenrm \scriptscriptfont0=\sixrm
\textfont1=\teni \scriptfont1=\seveni \scriptscriptfont1=\sixi
\textfont2=\tensy \scriptfont2=\sevensy \scriptscriptfont2=\sixsy
\textfont3=\tenex \scriptfont3=\tenex \scriptscriptfont3=\tenex
\textfont\itfam=\tenit \textfont\slfam=\tensl
\textfont\bffam=\tenbf \textfont\ttfam=\tentt
}

\def\twelvepoint{%
\def\rm{\fam0\twelverm}%
\def\it{\fam\itfam\twelveit}%
\def\sl{\fam\slfam\twelvesl}%
\def\bf{\fam\bffam\twelvebf}%
\def\tt{\fam\ttfam\twelvett}%
\def\cal{\twelvesy}%
 \textfont0=\twelverm \scriptfont0=\sevenrm \scriptscriptfont0=\sixrm
\textfont1=\twelvei \scriptfont1=\seveni \scriptscriptfont1=\sixi
\textfont2=\twelvesy \scriptfont2=\sevensy \scriptscriptfont2=\sixsy
\textfont3=\tenex \scriptfont3=\tenex \scriptscriptfont3=\tenex
\textfont\itfam=\twelveit \textfont\slfam=\twelvesl
\textfont\bffam=\twelvebf \textfont\ttfam=\twelvett
}


\twelvepoint\rm \baselineskip 6.0 mm
\parindent 18 pt \parskip 0 mm

\input epsf.sty

\def\va{\vskip 5 mm}  \def\vc{\vskip 1.5 mm}
\def\ha{\hskip 5 mm}  \def\hc{\hskip 1.5 mm}

\centerline{\fourteenbf 
Tables of the partition functions for nickel, Ni I -- Ni X. }
\vskip 12 mm
\centerline{\bf J. Halenka$^1$, J. Madej$^2$, K. Langer$^1$, 
   and A. Mamok$^1$ }
\midinsert\obeylines \baselineskip 4.0 mm \leftskip 27 mm \vskip 12 mm
 $^1$Institute of Physics, University of Opole,
\vc
 \hc  Oleska 48, 45--052 Opole, Poland 

\vskip 5 mm
 $^2$Astronomical Observatory, University of Warsaw,
\vc
 \hc  Al. Ujazdowskie 4, 00-478 Warszawa, Poland

\endinsert

\vskip 20 mm
\sectionhead{ABSTRACT}
\vskip 10 mm \baselineskip 6.0 mm

We present extensive tables of the atomic partition
function ($APF$) for nickel ions, Ni~I -- Ni X. Partition functions
are given over wide range of temperature, $10^3 \, {\rm K} < T < 10^6$ K,
and lowering of ionization energy ($0.001 \, {\rm eV}< LIE < 5.0 $ eV), 
both taken as independent variables. Our $APF$ take into account all energy 
levels predicted by quantum mechanics, including autoionization levels.
The tables can be applied for the computations of model 
stellar atmospheres and theoretical spectra over very wide range of
spectral classes, from the coolest K--M dwarfs up to the hottest main
sequence, giant, and white dwarf stars. This include also model spectra 
of supersoft X-ray sources and accretion discs in interacting binaries.

Our tables are available at
{\tt http://www.astrouw.edu.pl/$\sim$acta/acta.html}
(Acta Astronomica Archive),
and {\tt http://draco.uni.opole.pl/Halenka.html}.

\vskip 2 cm
\line{{\bf Key words:} Atomic data -- Plasmas -- Stars: atmospheres \hfil}

\endpage

\line{}
\sectionhead{1. Introduction}
\vskip 3 mm

The knowledge of realiable atomic partition functions ($APF$) is
of extreme importance for the determination of ionization balance
in astrophysical plasmas. In particular, partition functions are
coefficients of the Saha-Eggert equations describing ionization 
states of elements in stellar atmospheres (equation of state for
plasma in Local Thermodynamic Equilibrium). Therefore the knowledge 
of numerical values of the $APF$ is essential for computations of
model stellar atmospheres and theoretical spectra, and for correct
interpretation of intensities of absorption spectral lines and
abundance determination of elements in stellar atmospheres.
\newtoks\headline \headline={\hss\tenrm\folio\hss}

Given value of the partition function $U$ for a particular ion depends 
on gas temperature $T$, and the local electron concentration $N_e$. 
Temperature enters partition function by its definition (e.g. Griem
1964, Drawin and Felenbok 1965, Traving et al. 1966):

$$ U(T, N_e) = \sum\limits^{i_{max}}_{i=1} g_i \exp (-E_i / kT) \, , 
   \eqno(1) $$

\noindent
where the sum is taken over discrete energy levels of statistical weight
$i$ and excitation energies $E_i$. In vacuum ($N_e = 0$) the number of 
bound levels $i_{max}$ is infinite, and therefore the above series always
diverges.

In real plasma, however, interaction between the atom of interest and
surrounding free electrons and ions (plasma effects) cause, that bound
levels of very high excitation energies move to continuum and no longer
contribute to the partitiion function. Therefore the series in Eq. (1)
reduces to finite number of terms, and value of $U$ is also finite
and is strongly dependent on the electron concentration $N_e$. In
general, the larger is $N_e$ the lower is both number of bound energy
levels $i_{max}$ in Eq. (1), and the value of $U$.

There exist a large number of papers, which present tables of rather 
approximate (or even schematic) partition functions ($APF$) for elements,
including nickel ions (Drawin and Felenbok 1965; Traving et al. 1966; 
Irwin 1981, for example). The latter paper present fitting formulae for
$APF$ of Ni I -- Ni III, for temperatures $T \le 16000$ K. The widely
used computer code Tlusty 195 for computations of NLTE model stellar 
atmospheres (Hubeny and Lanz 1992, 1995) contains FORTRAN subroutine 
computing $APF$ of Ni IV -- Ni IX by direct summation over all 
{\sl observed} energy levels of these ions. However, all these
partition functions for Ni ions depend only on temperature $T$, and no
level dissolution with increasing density is included here. The latter
implies, that the set of energy levels was not complete there.

Quality of given tables of partition functions depend on ($i$)
accuracy and completeness of energy levels included in Eq. (1),
and ($ii$) realiability of the assumed theory of emitter-plasma
interactions. Emitter-plasma interactions cause that the series in 
Eq. (1) is finite. Unfortunately, none of the currently existing theories 
describe correctly effects of charged particles in plasma on the
atomic partition functions, cf. also Hummer and Mihalas (1988). 

Taking this into account we have decided to compute and present tables 
of the partition functions taking into account all energy
levels predicted by quantum mechanics, including also levels lying
above the so called {\sl normal ionization energy} (autoionization levels).
Moreover, our computations are more physically correct, since the $APF$
depend on both temperature $T$ and electron concentration $N_e$, through
tabulated values of lowering of the ionization energy ($LIE$). 
The method of $APF$ calculations is briefly described in the following 
Section. Detailed description of this method was presented in a series
of papers by Halenka and Grabowski (1977, 1984, 1986), Halenka (1988, 1989),
and Madej et al. (1999).

\sectionhead{2. Computation of the atomic partition functions}
\vskip 3 mm

We define the atomic partition function of $r$-th ionization state
by the equation

$$ U^ {(r)} (T, N_e) = \sum\limits_{p=1}^{p_{max}}
   \sum\limits_{i=1}^{i(p)_{max}} g^{(r)}_{pi} \exp(-E^{(r)}_{pi} /kT)
   = \sum\limits_{p=1}^{p_{max}} U^{(r)}_p (T, N_e) \, ,   \eqno(2) $$

\noindent
cf. Halenka and Grabowski (1977).
Here the set ($pi$) with the indices $p$ and $i$ (ordering of levels
from the ground level upwards in energy scale) describes an eigenstate 
of the atom in the $r$-th ionization state. Index $i$ represents three
quantum numbers ($nlj$) of the optical electron, and $p$ represents the
quantum state of the atomic core. Index $i(p)_{max}$ is the number of
all bound energy levels, $g^{(r)}_{pi}$ and $E^{(r)}_{pi}$ denote
statistical weight and excitation energy of the $i$-th state, in the
sequence based on the $p$-th parent level. Numbers $i(p)_{max}$ result
from the inequality

$$ E^{(r)}_{pi} \le E^{(r)}_{p\infty} - \Delta E^{(r)} \, ,  \eqno(3) $$

\noindent
where $\Delta E^{(r)}$ denotes the lowering of the ionization energy $LIE$,
and $E^{(r)}_{p\infty}$ is the ionization energy in the $p$-th level
sequence. The latter quantity is equal to the sum

$$ E^{(r)}_{p\infty} = E^{(r)}_{1\infty} + E^{(r+1)}_p \, .  \eqno(4)$$

\noindent
The quantity $ E^{(r+1)}_p$ denotes the energy of the atomic core after
ionization, $r \rightarrow r+1$.
Index $p_{max}$ is the number of different parent levels which can
be realized in given physical conditions. The number $p_{max}$ results
from the inequality similar to Eq. (3), written for the $(r+1)$-th 
ionization state. Since for a fixed value of $p$ the number $k$ is
assigned unambigously, then the upper limit of $E^{(r)}_{p\infty}$ for
the $k$-fold excitation ($k=1,2, \ldots, Z-r$, where $Z$ is the
atomic number) can be written as follows

$$ E^{(r)}_{p\infty} \le \sum\limits^k_{s=1} E^{(r+s-1)}_{1\infty} \, .
   \eqno(5) $$

Following Eq. (3) we have computed extensive tables of the $APF$ for nickel
in the 10 lowest ionization states, Ni I -- Ni X. 
Excitation energies and statistical weights of the ``observed'' levels 
available for nickel ions were taken from Kurucz (1994). We have added
to our partition functions contribution from many energy levels predicted
by quantum mechanics, but missed in his catalogue. This include also many
autoionizing levels. Exact description of the method of adding of missing 
levels is given in Halenka and Grabowski (1977, 1984).

In practical model atmosphere calculations one has to estimate the
lowering of ionization energy for nickel as function of temperature $T$ 
and electron concentration $N_e$. Such a relation has to result from
the model describing plasma-emitter interaction. Unfortunately, none of
the existing models is satisfactory enough (Hummer and Mihalas 1988).
We suggest use of the approximate relation between the $LIE$ and $N_e$
$$ {\delt} \chi = Ze^2 /D = 3 \times 10^{-8} Z N_e^{1/2} \,
     T^{-1/2} \hskip 6 mm \rm [eV]  \eqno(6) $$
(Eq. 9-106 of Mihalas, 1978), where $D= 4.8 \, (T/N_e)^{1/2}$ [cm] is
the Debye length in hydrogen dominated plasma. However, other relations
of this type are also given by Drawin and Felenbok (1965).

\sectionhead{3. Results}
\vskip 3 mm

As the ilustration of our recommended results, Fig. 1 presents run of 
$APF$ for Ni V and temperatures corresponding to atmospheres of hot white
dwarf stars, for various values of $LIE$ taken as free parameter (solid 
lines). Numerical values of $APF$ for Ni V are listed in Table 1.
For a comparison, we have computed also partition functions taking
into account so called {\sl observed} levels (single dashed line). One can
easily note, that our recommended partition functions are significantly 
larger than the latter functions. 

Partition functions for nickel are arranged in 10 ASCII tables, where
each table corresponds to a single ion. 
Complete set of these data is available from Acta Astronomica Archive ,
{\tt http://www.astrouw.edu.pl/$\sim$acta/acta.html}, \eol
or {\tt http://draco.uni.opole.pl/Halenka.html}.

Entries of a table are decimal 
logarithms of the APF. They are tabulated at 62 discrete temperatures,
spaced at nonequidistant intervals, $10^3 \le T \le 10^6$ K, and at 
9 arbitrarily assumed values of the lowering of ionization energy ($LIE$
= 0.001, 0.003, 0.010, 0.030, 0.100, 0.300, 1.000, 3.000, and 5.000 eV).
Both $T$ and $LIE$ points remain identical in all 10 tables of APF, to
ensure homogeneity of the data.

We are aware, that some values of $T$ and $LIE$ in our tables do not
correspond to conditions met in astrophysical plasma. However, the
extend of all tables ensures, that they cover practically all
$T$ and $LIE$ expected in stellar atmospheres of any type, excluding
atmospheres of known neutron stars (temperatures $T \ge 10^6$ K).

Tables presented in this paper are based on the most complete set of  
energy levels actually available. Moreover, our partition functions are
sensitive for plasma interactions, i.e. they strongly depend on the
lowering of ionization energy, which is expected in plasma. Our $APF$
tend to diverge for the lowering of ionization energy approaching zero,
which is the fundamental property of the partition functions in general.



\epsfxsize=15.0cm \epsfbox[60 80 630 575]{xac2.fig}

\leftskip 19 mm \rightskip 31 mm
\tenpoint\rm \baselineskip 4.5 mm

\vskip -30 mm
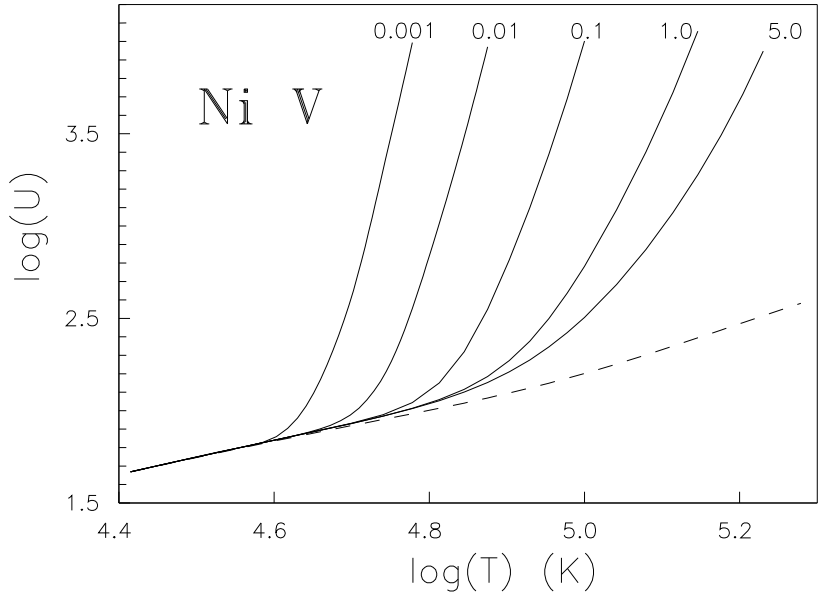
{\bf Fig. 1 :} Run of partition functions of Ni V as function of gas
temperature $T$ and various parameters $LIE$. Solid lines represent our
recommended results, whereas dashed line represents partition function
computed from the observed levels only.

\twelvepoint\rm \baselineskip 6.0 mm
\parindent 21 pt \parskip 0 mm
\leftskip 0 mm \rightskip 0 mm

\vskip 20 mm
\line{\bf Acknowledgements \hfil} \va  
We are grateful to R.L. Kurucz for making his CD-ROM No. 22 available
for us. JH and JM acknowledge support by grant No. 2 P03D 013 19 from the
Polish Committee for Scientific Research.

\endpage

\line{}
\sectionhead{REFERENCES}
\va
\dimen0=\hsize  \advance\dimen0 by -25 pt
\def\ref#1{\parshape=2  0.pt \hsize  25pt \dimen0 #1}

\parindent 0pt
\parskip 8pt
   
\ref Drawin, H.W., and Felenbok, P. 1965, Data for Plasma in Local
     Thermodynamic Equilibrium (Gauthier-Villars, Paris)

\ref Griem, H.R. 1964, Plasma Spectroscopy (McGraw-Hill Book Co.,
     New York)

\ref Halenka, J. 1988, Astron. Astrophys. Suppl., {\bf 75}, 47                            

\ref Halenka, J. 1989, Astron. Astrophys. Suppl., {\bf 81}, 303

\ref Halenka, J., and Grabowski, B. 1977, Astron. Astrophys., {\bf 54}, 757
             
\ref Halenka, J., and Grabowski, B. 1984, Astron. Astrophys. Suppl.,
     {\bf 57}, 43

\ref Halenka, J., and Grabowski, B. 1986, Astron. Astrophys. Suppl.,
     {\bf 64}, 495

\ref Hubeny, I., and Lanz, T. 1992, Astron. Astrophys., {\bf 262}, 501

\ref Hubeny, I., and Lanz, T. 1995, Astrophys. J, {\bf 439}, 875

\ref Hummer, D.G., and Mihalas, D. 1988, Astrophys. J., {\bf 331}, 794

\ref Irwin, A.W. 1981, Astrophys. J. Suppl., {\bf 45}, 621

\ref Kurucz, R.L. 1994, CD-ROM No. 22

\ref Madej, J., Halenka, J., and Grabowski, B. 1999, Astron. Astrophys.,
     {\bf 343}, 531

\ref Mihalas, D. 1978, Stellar Atmospheres (Freeman, San Francisco)

\ref Traving, G., Baschek, B., and Holweger, H. 1966, Abh. Hamburger
    Sternw., {\bf VIII}, No. 1

\parindent 21 pt
\endpage

\tenpoint\rm \baselineskip 3.5 mm

\newdimen\digitwidth
\setbox0=\hbox{\rm0}
\digitwidth=\wd0  
\catcode`?=\active
\def?{\kern\digitwidth}
\def\-{$-$}

\newtoks\headline \headline={\hfil}

\halign {\hfil#\hfil& \hskip 3 mm \hfil#\quad\hfil& \hfil#\quad\hfil
    & \hfil#\quad\hfil & \hfil#\quad\hfil & \hfil#\quad\hfil & \hfil#\quad\hfil
    & \hfil#\quad\hfil & \hfil#\quad\hfil & \hfil#\quad\hfil   \cr
\noalign{\centerline{Table 1} \smallskip}
\noalign{\centerline{Atomic partition functions for Ni V \ha 
   (decimal logarithms)} }
\noalign{\medskip \hrule \smallskip }
\noalign{\centerline{Lowering of ionization energy (eV)}  \smallskip }
  \ha T (K)& ??0.001& ??0.003& ??0.010& ??0.030& ??0.100& ??0.300& ??1.000& 
        ??3.000& ??5.000\cr
\noalign{\smallskip \hrule \smallskip}
???1000&?1.0714&?1.0714&?1.0714&?1.0714&?1.0714&?1.0714&?1.0714&?1.0714&?1.0714\cr
???3000&?1.2458&?1.2458&?1.2458&?1.2458&?1.2458&?1.2458&?1.2458&?1.2458&?1.2458\cr
???6000&?1.3173&?1.3173&?1.3173&?1.3173&?1.3173&?1.3173&?1.3173&?1.3173&?1.3173\cr
???8000&?1.3447&?1.3447&?1.3447&?1.3447&?1.3447&?1.3447&?1.3447&?1.3447&?1.3447\cr
??10000&?1.3740&?1.3740&?1.3740&?1.3740&?1.3740&?1.3740&?1.3740&?1.3740&?1.3740\cr
??12000&?1.4081&?1.4081&?1.4081&?1.4081&?1.4081&?1.4081&?1.4081&?1.4081&?1.4081\cr
??14000&?1.4462&?1.4462&?1.4462&?1.4462&?1.4462&?1.4462&?1.4462&?1.4462&?1.4462\cr
??16000&?1.4861&?1.4861&?1.4861&?1.4861&?1.4861&?1.4861&?1.4861&?1.4861&?1.4861\cr
??18000&?1.5261&?1.5261&?1.5261&?1.5261&?1.5261&?1.5261&?1.5261&?1.5261&?1.5261\cr
??20000&?1.5650&?1.5650&?1.5650&?1.5650&?1.5650&?1.5650&?1.5650&?1.5650&?1.5650\cr
??23000&?1.6196&?1.6196&?1.6196&?1.6196&?1.6196&?1.6196&?1.6196&?1.6196&?1.6196\cr 
??26000&?1.6692&?1.6692&?1.6692&?1.6692&?1.6692&?1.6692&?1.6692&?1.6692&?1.6692\cr 
??30000&?1.7274&?1.7274&?1.7274&?1.7274&?1.7274&?1.7274&?1.7274&?1.7274&?1.7274\cr 
??35000&?1.7898&?1.7894&?1.7893&?1.7893&?1.7893&?1.7893&?1.7893&?1.7893&?1.7893\cr 
??40000&?1.8575&?1.8448&?1.8422&?1.8418&?1.8417&?1.8417&?1.8417&?1.8417&?1.8417\cr
??50000&?2.6059&?2.1647&?1.9781&?1.9395&?1.9314&?1.9302&?1.9299&?1.9298&?1.9298\cr
??55000&?3.3019&?2.6546&?2.1811&?2.0193&?1.9787&?1.9721&?1.9708&?1.9704&?1.9703\cr
??60000&?3.9935&?3.2352&?2.5551&?2.1802&?2.0438&?2.0181&?2.0131&?2.0115&?2.0112\cr
??65000&?4.7515&?3.8183&?3.0138&?2.4548&?2.1501&?2.0756&?2.0601&?2.0551&?2.0543\cr
??70000&?5.6136&?4.4428&?3.4848&?2.8054&?2.3193&?2.1551&?2.1158&?2.1030&?2.1010\cr
??75000&?6.5202&?5.1417&?3.9712&?3.1802&?2.5493&?2.2667&?2.1848&?2.1571&?2.1527\cr
??80000&?7.4633&?5.8866&?4.4993&?3.5634&?2.8165&?2.4143&?2.2711&?2.2188&?2.2104\cr
??85000&?8.4515&?6.6585&?5.0713&?3.9642&?3.0992&?2.5924&?2.3763&?2.2890&?2.2745\cr
??90000&?9.4557&?7.4630&?5.6713&?4.3931&?3.3893&?2.7901&?2.4992&?2.3680&?2.3452\cr
??95000&10.4622&?8.2918&?6.2927&?4.8493&?3.6888&?2.9978&?2.6360&?2.4553&?2.4222\cr
?100000&11.4835&?9.1284&?6.9361&?5.3258&?4.0023&?3.2103&?2.7822&?2.5501&?2.5047\cr
?110000&13.5670&10.8261&?8.2657&?6.3273&?4.6774&?3.6474&?3.0883&?2.7565&?2.6829\cr
?120000&15.6782&12.5674&?9.6250&?7.3798&?5.4067&?4.1113&?3.4018&?2.9781&?2.8734\cr
?130000&17.7822&14.3359&11.0204&?8.4622&?6.1791&?4.6108&?3.7226&?3.2103&?3.0721\cr
?140000&19.8469&16.1070&12.4448&?9.5740&?6.9842&?5.1464&?4.0558&?3.4522&?3.2780\cr
?150000&21.8756&17.8536&13.8858&10.7152&?7.8155&?5.7149&?4.4051&?3.7054&?3.4916\cr
?160000&23.8485&19.5773&15.3207&11.8792&?8.6733&?6.3124&?4.7723&?3.9711&?3.7145\cr
?170000&25.7824&21.2664&16.7411&13.0523&?9.5563&?6.9356&?5.1574&?4.2504&?3.9479\cr
?180000&27.6557&22.9208&18.1430&14.2208&10.4593&?7.5829&?5.5597&?4.5434&?4.1927\cr
?190000&29.4117&24.5349&19.5193&15.3780&11.3738&?8.2518&?5.9778&?4.8495&?4.4490\cr
?200000&31.0247&26.0761&20.8675&16.5179&12.2911&?8.9376&?6.4101&?5.1677&?4.7162\cr
?220000&33.8432&28.8464&23.4343&18.7264&14.1099&10.3366&?7.3092&?5.8338&?5.2779\cr
?230000&35.0755&30.0715&24.6189&19.7872&15.0010&11.0389&?7.7710&?6.1777&?5.5689\cr
?240000&36.2077&31.2000&25.7258&20.8097&15.8734&11.7375&?8.2371&?6.5258&?5.8640\cr
?250000&37.2513&32.2417&26.7557&21.7855&16.7232&12.4292&?8.7053&?6.8760&?6.1615\cr
?265000&38.6719&33.6610&28.1660&23.1506&17.9478&13.4466&?9.4069&?7.4012&?6.6084\cr
?280000&39.9431&34.9316&29.4326&24.3949&19.1021&14.4319&10.1039&?7.9219&?7.0523\cr
?300000&41.4437&36.4318&30.9305&25.8789&20.5184&15.6831&11.0203&?8.6033&?7.6338\cr
?325000&43.0639&38.0519&32.5495&27.4908&22.0898&17.1285&12.1351&?9.4278&?8.3378\cr
?350000&44.4561&39.4440&33.9412&28.8797&23.4601&18.4309&13.2012&10.2180&?9.0128\cr
?375000&45.6654&40.6533&35.1503&30.0874&24.6586&19.5929&14.2040&10.9714&?9.6574\cr
?400000&46.7255&41.7134&36.2104&31.1469&25.7129&20.6271&15.1337&11.6861&10.2711\cr
?425000&47.6625&42.6505&37.1475&32.0837&26.6465&21.5491&15.9873&12.3606&10.8540\cr
?450000&48.4968&43.4847&37.9818&32.9178&27.4787&22.3743&16.7669&12.9940&11.4058\cr
?475000&49.2443&44.2323&38.7293&33.6652&28.2248&23.1160&17.4774&13.5861&11.9265\cr
?500000&49.9180&44.9060&39.4030&34.3389&28.8975&23.7859&18.1254&14.1380&12.4166\cr
?525000&50.5282&45.5163&40.0134&34.9492&29.5070&24.3936&18.7171&14.6513&12.8768\cr
?550000&51.0837&46.0717&40.5689&35.5047&30.0620&24.9472&19.2589&15.1282&13.3082\cr
?600000&52.0574&47.0455&41.5427&36.4785&31.0349&25.9187&20.2141&15.9831&14.0903\cr
?650000&52.8830&47.8711&42.3684&37.3043&31.8600&26.7430&21.0280&16.7235&14.7760\cr
?700000&53.5921&48.5802&43.0775&38.0134&32.5687&27.4513&21.7292&17.3683&15.3787\cr
?750000&54.2078&49.1959&43.6932&38.6292&33.1841&28.0666&22.3392&17.9336&15.9107\cr
?800000&54.7474&49.7355&44.2329&39.1689&33.7235&28.6059&22.8746&18.4325&16.3826\cr
?850000&55.2244&50.2125&44.7099&39.6459&34.2002&29.0826&23.3482&18.8756&16.8036\cr
?900000&55.6490&50.6371&45.1346&40.0706&34.6247&29.5071&23.7701&19.2717&17.1810\cr
?950000&56.0295&51.0176&45.5151&40.4511&35.0050&29.8875&24.1484&19.6277&17.5211\cr
1000000&56.3723&51.3605&45.8580&40.7941&35.3478&30.2303&24.4895&19.9493&17.8290\cr
\noalign{\medskip \hrule}  }

\par\vfill\end